\documentclass[twocolumn]{revtex4}
\usepackage{graphicx}

\begin{document}

\title{Strained single-crystal Al$_2$O$_3$ grown layer-by-layer on Nb (110) thin films\footnote{This article has been submitted to Applied Physics Letters.}}
\author{Paul B. Welander and James N. Eckstein}
\affiliation{Department of Physics and Frederick Seitz Materials Research Laboratory, University of Illinois at Urbana-Champaign, Urbana, IL 61801}
\date{April 1, 2007}
\begin{abstract}
We report on the layer-by-layer growth of single-crystal Al$_2$O$_3$ thin-films on Nb (110). Single-crystal Nb films are first prepared on A-plane sapphire, followed by the evaporation of Al in an O$_2$ background. The first stages of Al$_2$O$_3$ growth are layer-by-layer with hexagonal symmetry. Electron and x-ray diffraction measurements indicate the Al$_2$O$_3$ initially grows clamped to the Nb lattice with a tensile strain near $10\%$. This strain relaxes with further deposition, and beyond about 50 {\AA} we observe the onset of island growth. Despite the asymmetric misfit between the Al$_2$O$_3$ film and the Nb under-layer, the observed strain is surprisingly isotropic.
\end{abstract}
\maketitle

The present challenge of constructing solid-state quantum bits with long coherence times \cite{Nielson2000} has ignited new interest in Josephson junctions fabricated from single-crystal materials.  It has been found that critical-current $1/f$ noise cannot fully account for the observed decoherence times in junctions-based qubits \cite{vanHarlingen2004}.  However, amorphous tunnel-barrier defects can give rise to two-level \textit{charge} fluctuations that destroy quantum coherence across the junction \cite{Martin2005, Martinis2005}.  Oh \textit{et al} have recently found that tunnel-junctions from epitaxial Re/Al$_2$O$_3$/Al tri-layers have a significantly reduced density of two-level fluctuators \cite{Oh2006a}.

The pairing of Re and Al$_2$O$_3$ is advantageous because of the very small misfit between the basal planes and because Re is less likely to oxidize compared with other superconducting refractory metals.  However, epitaxial Re films develop domains due to basal-plane twinning, causing the surface to be rough on the length scales of a typical tunnel-junction \cite{Oh2006b}.  An alternative to Al$_2$O$_3$ hetero-epitaxy on a close-packed metal surface is to grow on bcc (110), where such twinning is absent.  To date single-crystal Al$_2$O$_3$ films have been grown on a number of such metals:  Ta \cite{Chen1994}, Mo \cite{Wu1994}, W \cite{Gunster1995}, and more recently Nb \cite{Dietrich2005}.

In a recent paper, Dietrich et al reported on their investigations of ultra-thin epitaxial $\alpha$-Al$_2$O$_3$ (0001) films on Nb using tunneling microscopy and spectroscopy \cite{Dietrich2005}.  Their films were grown on Nb (110) by evaporating Al in an O$_2$ background near room temperature.  Crystallization was achieved by annealing the sample up to 1000 $^{\circ}$C.  Subsequent microscopy showed the film to be atomically smooth, but spectroscopic scans found localized defect states around $\pm1$ eV, well below the 9 eV sapphire band gap.

We report here on our findings concerning the hetero-epitaxy of Al$_2$O$_3$ on Nb (110) films.  Unlike the previous study, our Al$_2$O$_3$ films are grown layer-by-layer with co-deposition of Al and O at elevated substrate temperatures.  Epitaxial bi-layers (Nb/Al$_2$O$_3$) and tri-layers (Nb/Al$_2$O$_3$/Nb) are grown by molecular beam epitaxy (MBE). Characterization techniques include \textit{in situ} reflection high-energy electron diffraction (RHEED) and x-ray photo-electron spectroscopy (XPS), and \textit{ex situ} atomic force microscopy (AFM) and x-ray diffraction (XRD).

The process for growing high-quality single-crystal Nb films on sapphire is well understood \cite{Durbin1982}.  Our samples start with a thick Nb base layer (2000 {\AA}) grown on A-plane sapphire -- $\alpha$-Al$_2$O$_3$ $(11\bar{2}0)$ -- with a nominal miscut of 0.1$^{\circ}$.  Nb ($99.99\%$) is evaporated via e-beam bombardment at a rate of about 0.3 {\AA}/s onto a substrate held near 800 $^{\circ}$C.  The base pressure of our chamber is about $10^{-11}$ torr, with the growth pressure around $10^{-9}$ torr.  After deposition, the film is annealed above 1300 $^{\circ}$C for 30 min.  During growth and annealing the film surface is monitored with RHEED.

\begin{figure}[b]
  \includegraphics[width=3.375in]{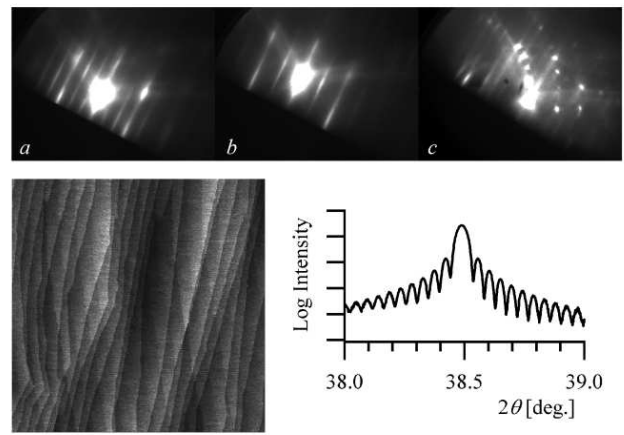}
  \caption{Nb (110) on A-plane sapphire.  Top: RHEED images along the (a) [001] and (b) $[1\bar{1}1]$ azimuths after growth at 800 $^{\circ}$C, and (c) $[1\bar{1}1]$ after annealing above 1300 $^{\circ}$C.  Left: $5\times5$ $\mu$m$^2$ AFM image of annealed Nb, 10 {\AA} height scale.  Right: XRD radial scan of the Nb (110) Bragg peak.}
  \label{Figure1}
\end{figure}

Epitaxial Nb on A-plane sapphire grows in the (110) orientation with Nb $[1\bar{1}1]$ $\parallel$ $\alpha$-Al$_2$O$_3$ [0001], in accordance with the well-established three-dimensional relationship \cite{Durbin1982, Mayer1990}.  Nb RHEED patterns (Figure \ref{Figure1}) reveal a two-dimensional, reconstructed film surface that takes one form after growth \cite{Surgers1992}, and a second one upon annealing \cite{Ondrejcek2001}.  Annealed films also show a sharp specular spot indicating long-range film flatness, which is confirmed by AFM measurements.  Scans show large terraces about 2000 {\AA} wide and monolayer step-edges that align themselves according to the substrate miscut (Figure \ref{Figure1}).  Annealed Nb films typically have an rms surface roughness less than 2 {\AA}.

XRD measurements on these Nb films show sharp Bragg peaks and narrow rocking curves, both indicative of single-crystal growth.  Figure \ref{Figure1} shows a radial scan ($2\theta$-$\omega$) of the Nb (110) Bragg peak from a 2000 {\AA}-thick film, with intensity fringes indicating a structural coherence that extends over the entire film thickness.  Rocking curves typically have a FWHM of about $0.03^{\circ}$.  In addition,  measurements of specular and off-axis Bragg peaks demonstrate that a 2000 {\AA}-thick annealed Nb film is strained $0.1\%$ or less with respect to bulk.

\begin{figure}
  \includegraphics[width=3.375in]{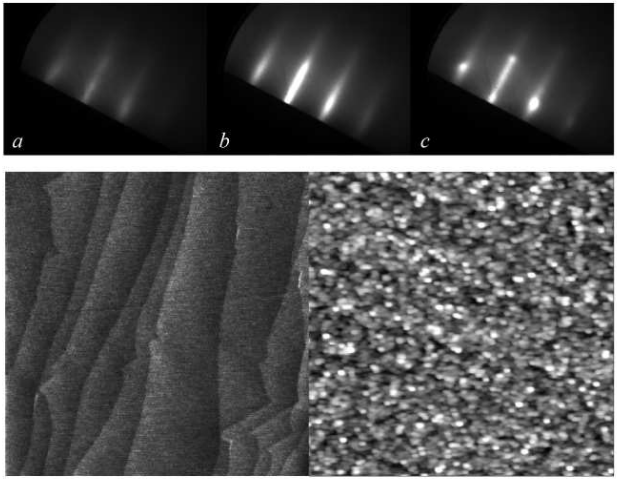}
  \caption{Top: RHEED images from epitaxial Al$_2$O$_3$ on Nb (110), taken along the $[1\bar{1}00]$ azimuth after deposition of (a) 4 {\AA}, (b) 25 {\AA}, and (c) 125 {\AA}.  Bottom: $5\times5$ $\mu$m$^2$ AFM scans on Al$_2$O$_3$ films that are 20 {\AA} (left, 10 {\AA} height-scale) and 100 {\AA} (right, 50 {\AA}) thick.}
  \label{Figure2}
\end{figure}

Al$_2$O$_3$ is deposited \textit{in situ} onto similar Nb films at a substrate temperature of around 750 $^{\circ}$C.  Using a standard effusion cell, Al ($99.9995\%$) is evaporated at about 0.1 {\AA}/s in an O$_2$ ($99.995\%$) background up to $5\times10^{-6}$ torr.  Under these growth conditions we estimate that the O$_2$ flux is about 1000 times greater than that of Al \cite{Tscherich1998}.  After deposition the sample is cooled before turning the O$_2$ off.  Al$_2$O$_3$ films included in this report range in thickness from 15 to 125 {\AA}.

Chemical analysis of the Al$_2$O$_3$ is carried out in an XPS system adjacent to the growth chamber.  Measurements of the Al $2p$, O $1s$ and Nb $3d$ levels indicate that the Al is completely oxidized with no measurable oxidation of the underlying Nb.  The observed energy difference between the O $1s$ and Al $2p$ levels is 457.1 eV, in good agreement with what has been reported for sapphire (456.6 eV) \cite{Mullins1988}.  The Nb $3d$ level shows no side bands which would indicate oxide formation.

RHEED of the Al$_2$O$_3$ thin film reveals a hexagonal C-plane-like surface in the Nishiyama-Wasserman orientation: $\alpha$-Al$_2$O$_3$ (0001) $[\bar{1}100]$ $\parallel$ Nb (110) [001] \cite{Bruce1978}.  (Because both $\alpha$-Al$_2$O$_3$ \cite{Lee1985} and $\gamma$-Al$_2$O$_3$ \cite{Streitz1999} have close-packed planes, no definitive crystal structure can be inferred.  Hexagonal Miller indices will be employed for defining crystallographic orientations by convention only.)  Diffraction images from various stages of growth are shown in Figure \ref{Figure2}.  Immediately after the oxide deposition begins the Nb diffraction pattern and specular spot disappear.  After about 2 ML (4 {\AA}) the Al$_2$O$_3$ diffraction pattern becomes visible.  At a thickness of 25 {\AA}, RHEED shows an elongated specular spot and well-defined first-order streaks.  Up to about 50 {\AA} the Al$_2$O$_3$ growth is layer-by-layer (Frank-van der Merwe mode).  Beyond this thickness the 2D streaks evolve into 3D spots, indicating the growth of islands (Stranski-Krastanov mode).

\begin{figure}
  \includegraphics[width=3.375in]{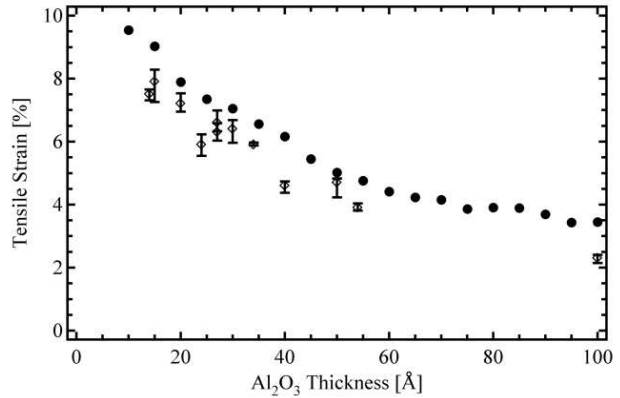}
  \caption{Strain vs. film thickness for epitaxial Al$_2$O$_3$ on Nb (110).  ($\bullet$) Strain of a 100 {\AA} film measured during deposition.  ($\diamond$) Strain observed for a number of samples after deposition and cooling, with error bars indicating the range of strain values measured along different RHEED azimuths.}
  \label{Figure3}
\end{figure}

As the transformation from 2D to 3D growth is occurring, the measured spacing between RHEED streaks/spots increases, indicating a shrinking of the Al$_2$O$_3$ surface lattice.  Using the RHEED from the base-layer Nb as a ruler, we find that the Al$_2$O$_3$ film experiences a tensile strain that relaxes with increasing thickness, as shown in Figure \ref{Figure3}.  The strain-thickness curve is determined from RHEED along the $[\bar{1}100]$ azimuth during Al$_2$O$_3$ deposition near 750 $^{\circ}$C.  With respect to C-plane sapphire ($a$ = 4.759 {\AA}), the tensile strain is nearly $10\%$ initially and by 20 {\AA} has fallen to about $8\%$.  After 100 {\AA} of deposition, the Al$_2$O$_3$ exhibits a tensile strain of around $3\%$.

After deposition and cooling in O$_2$, Al$_2$O$_3$ films of various thicknesses show further lattice relaxation (Figure \ref{Figure3}).  On average, RHEED measurements near room temperature show a strain reduction of about $1\%$ when compared to measurements just after Al deposition.  Thermal contraction accounts for a significant portion of the strain change during cooling.  (Both Nb and Al$_2$O$_3$ have expansion coefficients in this temperature range around 7-8$\times10^{-6}$ K$^{-1}$.)  However, due to the limited precision of our measurements, the presence of other strain-relief mechanisms cannot be determined.

Regardless, the measured tensile strain in epitaxial Al$_2$O$_3$ films on Nb (110) is significant.  What's more, the strain is fairly isotropic -- RHEED patterns along the $\{\bar{1}100\}$ azimuths reveal relatively small variations.  The strain for each azimuth is determined by averaging opposite directions -- eg. $[\bar{1}100]$ and $[1\bar{1}00]$ -- to reduce systematic errors.  The mean and range of the measured tensile strain for the three azimuths is shown in Figure 3.  The strain-isotropy is surprising since the misfit along the Nb [001] or $\alpha$-Al$_2$O$_3$ $[\bar{1}100]$ direction is rather large ($20\%$), while along the Nb $[1\bar{1}0]$ or $\alpha$-Al$_2$O$_3$ $[\bar{1}\bar{1}20]$ it is much smaller ($-1.7\%$).  Despite such an anisotropic misfit, the Al$_2$O$_3$ films exhibit isotropic strain.

Thin Al$_2$O$_3$ films are also very flat.  AFM imaging of a 20 {\AA}-thick film shows an atomically flat surface with monolayer steps (c/6 = 2.165 {\AA}) and an rms roughness of about 2 {\AA} (Figure \ref{Figure2}).  On the other hand, the surface of a 100 {\AA}-thick film is comprised of islands about 1000 {\AA} wide and 50 {\AA} in height.  This agrees well with our interpretation of Al$_2$O$_3$ RHEED - evidence for islands in the diffraction images appeared after about 50 {\AA} of deposition.

\begin{figure}[t]
  \includegraphics[width=3.375in]{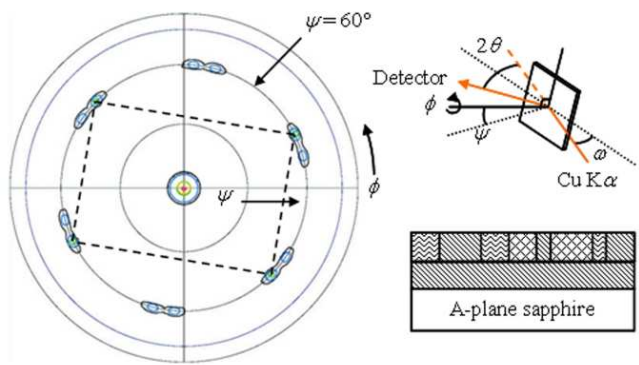}
  \caption{XRD pole figure for an epitaxial Nb/Al$_2$O$_3$/Nb tri-layer grown on A-plane sapphire.  Both Nb layers have a (110) surface-orientation.  This scan shows the off-axis $\langle110\rangle$ Bragg peaks.  The four peaks connected by the dashed rectangle are approximately four times stronger than the others.}
  \label{Figure4}
\end{figure}

For those samples where an epitaxial Nb over-layer is deposited \textit{in situ}, the substrate is warmed back up above 700 $^{\circ}$C.  Under these conditions growth on C-plane sapphire would yield (111)-oriented films \cite{Durbin1982, Mayer1990}.  However, XRD analysis indicates that the top Nb layer is (110)-oriented with Nb [001] $\parallel$ $\alpha$-Al$_2$O$_3$ $[\bar{1}100]$, $[0\bar{1}10]$ and $[10\bar{1}0]$.  A pole scan of off-axis $\langle110\rangle$ Bragg peaks is shown in Figure \ref{Figure4}, and despite the surface orientation, the Nb over-layer reproduces the hexagonal symmetry of the Al$_2$O$_3$ film.  The top Nb film grows in three domains of roughly equal weight rotated with respect to one another by 120$^{\circ}$, with one domain aligned to the base Nb layer.  This type of film structure has been observed for Nb growth on C-plane sapphire, but only under the following conditions: evaporation above 1000 $^{\circ}$C \cite{Wagner1998}, post-growth annealing up to 1500 $^{\circ}$C \cite{Wagner1996}, and niobium sputtering near 850 $^{\circ}$C \cite{Dietrich2003}.  That we observe this growth structure for evaporation near 700 $^{\circ}$C suggests that the surface lattice of the Al$_2$O$_3$ film, while hexagonal, is not identical to that of C-plane sapphire.

Tunnel-junctions were fabricated from several of these epitaxial tri-layers.  The $I$-$V$ characteristics showed a large conductance shunting the Josephson junction.  While an inhomogeneous morphology may cause such a conductance, no metallurgical pinholes were ever observed in our Al$_2$O$_3$ films.  Devices with 20 {\AA} Al$_2$O$_3$ layers had critical current densities around 10$^4$ A/cm$^2$ and normal state conductances near 10$^9$ S/cm$^2$.  Assuming a homogeneous barrier, the latter value gives an effective barrier height of about 1.3 eV.  This is similar to the energy of sub-gap states found spectroscopically by Dietrich \textit{et al} \cite{Dietrich2005} in epitaxial Al$_2$O$_3$ on Nb.

Among the previous studies of Al$_2$O$_3$ epitaxy on bcc (110) metals, only Chen \textit{et al} reported any measure of tensile strain \cite{Chen1994}.  For Al$_2$O$_3$ films 5-40 {\AA} thick on Ta (110) they measured a lattice enlargement of about $9\%$.  The agreement with our findings could be expected since the lattice constants of Ta and Nb are nearly identical.  One difference though is that Chen \textit{et al} observed a Kurdjumov-Sachs relationship, $\alpha$-Al$_2$O$_3$ (0001) $[\bar{1}100]$ $\parallel$ Nb (110) $[1\bar{1}1]$ \cite{Chen1994}, instead of the Nishiyama-Wasserman orientation we observe.

In summary, single-crystal Nb/Al$_2$O$_3$ and Nb/Al$_2$O$_3$/Nb multi-layers were grown by MBE.  Various methods of materials analysis suggest these layers are all high-quality.  Our principal finding is that epitaxial Al$_2$O$_3$ on Nb (110) grows under uniform tensile strain, despite the anisotropic misfit.  As the Al$_2$O$_3$ film thickness is increased the strain relaxes and the surface roughens.  The over-layer Nb grows with a (110) surface orientation under growth conditions that would yield Nb (111) on C-plane sapphire.

AFM and XRD analysis was carried out in the Center for Microanalysis of Materials, University of Illinois at Urbana-Champaign, which is partially supported by the U.S. Department of Energy under grant DEFG02-91ER45439.  This project was funded by the National Science Foundation through grant EIA 01-21568.

\bibliography{alumina}

\begin{thebibliography}{22}
\expandafter\ifx\csname natexlab\endcsname\relax\def\natexlab#1{#1}\fi
\expandafter\ifx\csname bibnamefont\endcsname\relax
  \def\bibnamefont#1{#1}\fi
\expandafter\ifx\csname bibfnamefont\endcsname\relax
  \def\bibfnamefont#1{#1}\fi
\expandafter\ifx\csname citenamefont\endcsname\relax
  \def\citenamefont#1{#1}\fi
\expandafter\ifx\csname url\endcsname\relax
  \def\url#1{\texttt{#1}}\fi
\expandafter\ifx\csname urlprefix\endcsname\relax\def\urlprefix{URL }\fi
\providecommand{\bibinfo}[2]{#2}
\providecommand{\eprint}[2][]{\url{#2}}

\bibitem[{\citenamefont{Nielson and Chuang}(2000)}]{Nielson2000}
\bibinfo{author}{\bibfnamefont{M.~A.} \bibnamefont{Nielson}} \bibnamefont{and}
  \bibinfo{author}{\bibfnamefont{I.~L.} \bibnamefont{Chuang}},
  \emph{\bibinfo{title}{Quantum Computation and Quantum Information}}
  (\bibinfo{publisher}{Cambridge University Press}, \bibinfo{year}{2000}).

\bibitem[{\citenamefont{van Harlingen et~al.}(2004)\citenamefont{van Harlingen,
  Robertson, Plourde, Reichardt, Crane, and Clarke}}]{vanHarlingen2004}
\bibinfo{author}{\bibfnamefont{D.~J.} \bibnamefont{van Harlingen}},
  \bibinfo{author}{\bibfnamefont{T.~L.} \bibnamefont{Robertson}},
  \bibinfo{author}{\bibfnamefont{B.~L.~T.} \bibnamefont{Plourde}},
  \bibinfo{author}{\bibfnamefont{P.~A.} \bibnamefont{Reichardt}},
  \bibinfo{author}{\bibfnamefont{T.~A.} \bibnamefont{Crane}}, \bibnamefont{and}
  \bibinfo{author}{\bibfnamefont{J.}~\bibnamefont{Clarke}},
  \bibinfo{journal}{Phys. Rev. B} \textbf{\bibinfo{volume}{70}},
  \bibinfo{pages}{064517} (\bibinfo{year}{2004}).

\bibitem[{\citenamefont{Martin et~al.}(2005)\citenamefont{Martin, Bulaevskii,
  and Shnirman}}]{Martin2005}
\bibinfo{author}{\bibfnamefont{I.}~\bibnamefont{Martin}},
  \bibinfo{author}{\bibfnamefont{L.}~\bibnamefont{Bulaevskii}},
  \bibnamefont{and} \bibinfo{author}{\bibfnamefont{A.}~\bibnamefont{Shnirman}},
  \bibinfo{journal}{Phys. Rev. Lett.} \textbf{\bibinfo{volume}{95}},
  \bibinfo{pages}{127002} (\bibinfo{year}{2005}).

\bibitem[{\citenamefont{Martinis et~al.}(2005)\citenamefont{Martinis, Cooper,
  McDermott, Steffen, Ansmann, Osborn, Cicak, Oh, Pappas, Simmonds
  et~al.}}]{Martinis2005}
\bibinfo{author}{\bibfnamefont{J.~M.} \bibnamefont{Martinis}},
  \bibinfo{author}{\bibfnamefont{K.~B.} \bibnamefont{Cooper}},
  \bibinfo{author}{\bibfnamefont{R.}~\bibnamefont{McDermott}},
  \bibinfo{author}{\bibfnamefont{M.}~\bibnamefont{Steffen}},
  \bibinfo{author}{\bibfnamefont{M.}~\bibnamefont{Ansmann}},
  \bibinfo{author}{\bibfnamefont{K.~D.} \bibnamefont{Osborn}},
  \bibinfo{author}{\bibfnamefont{K.}~\bibnamefont{Cicak}},
  \bibinfo{author}{\bibfnamefont{S.}~\bibnamefont{Oh}},
  \bibinfo{author}{\bibfnamefont{D.~P.} \bibnamefont{Pappas}},
  \bibinfo{author}{\bibfnamefont{R.}~\bibnamefont{Simmonds}},
  \bibnamefont{et~al.}, \bibinfo{journal}{Phys. Rev. Lett.}
  \textbf{\bibinfo{volume}{95}}, \bibinfo{pages}{210503}
  (\bibinfo{year}{2005}).

\bibitem[{\citenamefont{Oh et~al.}(2006{\natexlab{a}})\citenamefont{Oh, Cicak,
  Kline, Sillanp\"{a}\"{a}, Osborn, Whittaker, Simmonds, and Pappas}}]{Oh2006a}
\bibinfo{author}{\bibfnamefont{S.}~\bibnamefont{Oh}},
  \bibinfo{author}{\bibfnamefont{K.}~\bibnamefont{Cicak}},
  \bibinfo{author}{\bibfnamefont{J.~S.} \bibnamefont{Kline}},
  \bibinfo{author}{\bibfnamefont{M.~A.} \bibnamefont{Sillanp\"{a}\"{a}}},
  \bibinfo{author}{\bibfnamefont{K.~D.} \bibnamefont{Osborn}},
  \bibinfo{author}{\bibfnamefont{J.~D.} \bibnamefont{Whittaker}},
  \bibinfo{author}{\bibfnamefont{R.~W.} \bibnamefont{Simmonds}},
  \bibnamefont{and} \bibinfo{author}{\bibfnamefont{D.~P.}
  \bibnamefont{Pappas}}, \bibinfo{journal}{Phys. Rev. B}
  \textbf{\bibinfo{volume}{74}}, \bibinfo{pages}{100502}
  (\bibinfo{year}{2006}{\natexlab{a}}).

\bibitem[{\citenamefont{Oh et~al.}(2006{\natexlab{b}})\citenamefont{Oh, Hite,
  Cicak, Osborn, Simmonds, McDermott, Cooper, Steffen, Martinis, and
  Pappas}}]{Oh2006b}
\bibinfo{author}{\bibfnamefont{S.}~\bibnamefont{Oh}},
  \bibinfo{author}{\bibfnamefont{D.~A.} \bibnamefont{Hite}},
  \bibinfo{author}{\bibfnamefont{K.}~\bibnamefont{Cicak}},
  \bibinfo{author}{\bibfnamefont{K.~D.} \bibnamefont{Osborn}},
  \bibinfo{author}{\bibfnamefont{R.~W.} \bibnamefont{Simmonds}},
  \bibinfo{author}{\bibfnamefont{R.}~\bibnamefont{McDermott}},
  \bibinfo{author}{\bibfnamefont{K.~B.} \bibnamefont{Cooper}},
  \bibinfo{author}{\bibfnamefont{M.}~\bibnamefont{Steffen}},
  \bibinfo{author}{\bibfnamefont{J.~M.} \bibnamefont{Martinis}},
  \bibnamefont{and} \bibinfo{author}{\bibfnamefont{D.~P.}
  \bibnamefont{Pappas}}, \bibinfo{journal}{Thin Solid Films}
  \textbf{\bibinfo{volume}{389}}, \bibinfo{pages}{496}
  (\bibinfo{year}{2006}{\natexlab{b}}).

\bibitem[{\citenamefont{Chen and Goodman}(1994)}]{Chen1994}
\bibinfo{author}{\bibfnamefont{P.~J.} \bibnamefont{Chen}} \bibnamefont{and}
  \bibinfo{author}{\bibfnamefont{D.~W.} \bibnamefont{Goodman}},
  \bibinfo{journal}{Surf. Sci. Lett.} \textbf{\bibinfo{volume}{312}},
  \bibinfo{pages}{L767} (\bibinfo{year}{1994}).

\bibitem[{\citenamefont{Wu and Goodman}(1994)}]{Wu1994}
\bibinfo{author}{\bibfnamefont{M.-C.} \bibnamefont{Wu}} \bibnamefont{and}
  \bibinfo{author}{\bibfnamefont{D.~W.} \bibnamefont{Goodman}},
  \bibinfo{journal}{J. Phys. Chem.} \textbf{\bibinfo{volume}{98}},
  \bibinfo{pages}{9874} (\bibinfo{year}{1994}).

\bibitem[{\citenamefont{G\"{u}nster et~al.}(1995)\citenamefont{G\"{u}nster,
  Brause, Mayer, Hitzke, and Kempter}}]{Gunster1995}
\bibinfo{author}{\bibfnamefont{J.}~\bibnamefont{G\"{u}nster}},
  \bibinfo{author}{\bibfnamefont{M.}~\bibnamefont{Brause}},
  \bibinfo{author}{\bibfnamefont{T.}~\bibnamefont{Mayer}},
  \bibinfo{author}{\bibfnamefont{A.}~\bibnamefont{Hitzke}}, \bibnamefont{and}
  \bibinfo{author}{\bibfnamefont{V.}~\bibnamefont{Kempter}},
  \bibinfo{journal}{Nuc. Instr. and Meth. in Phys. Res. B}
  \textbf{\bibinfo{volume}{100}}, \bibinfo{pages}{411} (\bibinfo{year}{1995}).

\bibitem[{\citenamefont{Dietrich et~al.}(2005)\citenamefont{Dietrich,
  Koslowski, and Ziemann}}]{Dietrich2005}
\bibinfo{author}{\bibfnamefont{C.}~\bibnamefont{Dietrich}},
  \bibinfo{author}{\bibfnamefont{B.}~\bibnamefont{Koslowski}},
  \bibnamefont{and} \bibinfo{author}{\bibfnamefont{P.}~\bibnamefont{Ziemann}},
  \bibinfo{journal}{J. Appl. Phys.} \textbf{\bibinfo{volume}{97}},
  \bibinfo{pages}{083515} (\bibinfo{year}{2005}).

\bibitem[{\citenamefont{Durbin et~al.}(1981)\citenamefont{Durbin, Cunningham,
  Mochel, and Flynn}}]{Durbin1982}
\bibinfo{author}{\bibfnamefont{S.~M.} \bibnamefont{Durbin}},
  \bibinfo{author}{\bibfnamefont{J.~E.} \bibnamefont{Cunningham}},
  \bibinfo{author}{\bibfnamefont{M.~E.} \bibnamefont{Mochel}},
  \bibnamefont{and} \bibinfo{author}{\bibfnamefont{C.~P.} \bibnamefont{Flynn}},
  \bibinfo{journal}{J. Phys. F: Met. Phys.} \textbf{\bibinfo{volume}{11}},
  \bibinfo{pages}{L223} (\bibinfo{year}{1981}).

\bibitem[{\citenamefont{Mayer et~al.}(1990)\citenamefont{Mayer, Flynn, and
  R\"{u}hle}}]{Mayer1990}
\bibinfo{author}{\bibfnamefont{J.}~\bibnamefont{Mayer}},
  \bibinfo{author}{\bibfnamefont{C.~P.} \bibnamefont{Flynn}}, \bibnamefont{and}
  \bibinfo{author}{\bibfnamefont{M.}~\bibnamefont{R\"{u}hle}},
  \bibinfo{journal}{Ultramicroscopy} \textbf{\bibinfo{volume}{33}},
  \bibinfo{pages}{51} (\bibinfo{year}{1990}).

\bibitem[{\citenamefont{S\"{u}rgers and v.~L\"{o}hneysen}(1992)}]{Surgers1992}
\bibinfo{author}{\bibfnamefont{C.}~\bibnamefont{S\"{u}rgers}} \bibnamefont{and}
  \bibinfo{author}{\bibfnamefont{H.}~\bibnamefont{v.~L\"{o}hneysen}},
  \bibinfo{journal}{Appl. Phys. A} \textbf{\bibinfo{volume}{54}},
  \bibinfo{pages}{350} (\bibinfo{year}{1992}).

\bibitem[{\citenamefont{Ondrejcek et~al.}(2001)\citenamefont{Ondrejcek,
  Appleton, Swiech, Petrova, and Flynn}}]{Ondrejcek2001}
\bibinfo{author}{\bibfnamefont{M.}~\bibnamefont{Ondrejcek}},
  \bibinfo{author}{\bibfnamefont{R.~S.} \bibnamefont{Appleton}},
  \bibinfo{author}{\bibfnamefont{W.}~\bibnamefont{Swiech}},
  \bibinfo{author}{\bibfnamefont{V.~L.} \bibnamefont{Petrova}},
  \bibnamefont{and} \bibinfo{author}{\bibfnamefont{C.~P.} \bibnamefont{Flynn}},
  \bibinfo{journal}{Phys. Rev. Lett.} \textbf{\bibinfo{volume}{87}},
  \bibinfo{pages}{116102} (\bibinfo{year}{2001}).

\bibitem[{\citenamefont{Tscherich and von Bonin}(1998)}]{Tscherich1998}
\bibinfo{author}{\bibfnamefont{K.~G.} \bibnamefont{Tscherich}}
  \bibnamefont{and} \bibinfo{author}{\bibfnamefont{V.}~\bibnamefont{von
  Bonin}}, \bibinfo{journal}{J. Appl. Phys.} \textbf{\bibinfo{volume}{84}},
  \bibinfo{pages}{4065} (\bibinfo{year}{1998}).

\bibitem[{\citenamefont{Mullins and Averbach}(1988)}]{Mullins1988}
\bibinfo{author}{\bibfnamefont{W.~M.} \bibnamefont{Mullins}} \bibnamefont{and}
  \bibinfo{author}{\bibfnamefont{B.~L.} \bibnamefont{Averbach}},
  \bibinfo{journal}{Surf. Sci.} \textbf{\bibinfo{volume}{206}},
  \bibinfo{pages}{29} (\bibinfo{year}{1988}).

\bibitem[{\citenamefont{Bruce and Jaeger}(1978)}]{Bruce1978}
\bibinfo{author}{\bibfnamefont{L.~A.} \bibnamefont{Bruce}} \bibnamefont{and}
  \bibinfo{author}{\bibfnamefont{H.}~\bibnamefont{Jaeger}},
  \bibinfo{journal}{Phil. Mag. A} \textbf{\bibinfo{volume}{38}},
  \bibinfo{pages}{223} (\bibinfo{year}{1978}).

\bibitem[{\citenamefont{Lee and Lagerlof}(1985)}]{Lee1985}
\bibinfo{author}{\bibfnamefont{W.~E.} \bibnamefont{Lee}} \bibnamefont{and}
  \bibinfo{author}{\bibfnamefont{K.~P.~D.} \bibnamefont{Lagerlof}},
  \bibinfo{journal}{J. Elec. Micros. Tech.} \textbf{\bibinfo{volume}{2}},
  \bibinfo{pages}{247} (\bibinfo{year}{1985}).

\bibitem[{\citenamefont{Streitz and Mintmire}(1999)}]{Streitz1999}
\bibinfo{author}{\bibfnamefont{F.~H.} \bibnamefont{Streitz}} \bibnamefont{and}
  \bibinfo{author}{\bibfnamefont{J.~W.} \bibnamefont{Mintmire}},
  \bibinfo{journal}{Phys. Rev. B} \textbf{\bibinfo{volume}{60}},
  \bibinfo{pages}{773} (\bibinfo{year}{1999}).

\bibitem[{\citenamefont{Wagner}(1998)}]{Wagner1998}
\bibinfo{author}{\bibfnamefont{T.}~\bibnamefont{Wagner}}, \bibinfo{journal}{J.
  Mater. Res.} \textbf{\bibinfo{volume}{13}}, \bibinfo{pages}{693}
  (\bibinfo{year}{1998}).

\bibitem[{\citenamefont{Wagner et~al.}(1996)\citenamefont{Wagner, Lorenz, and
  R\"{u}hle}}]{Wagner1996}
\bibinfo{author}{\bibfnamefont{T.}~\bibnamefont{Wagner}},
  \bibinfo{author}{\bibfnamefont{M.}~\bibnamefont{Lorenz}}, \bibnamefont{and}
  \bibinfo{author}{\bibfnamefont{M.}~\bibnamefont{R\"{u}hle}},
  \bibinfo{journal}{J. Mater. Res.} \textbf{\bibinfo{volume}{11}},
  \bibinfo{pages}{1255} (\bibinfo{year}{1996}).

\bibitem[{\citenamefont{Ch.~Dietrich and Koslowski}(2003)}]{Dietrich2003}
\bibinfo{author}{\bibfnamefont{H.-G.~B.} \bibnamefont{Ch.~Dietrich}}
  \bibnamefont{and}
  \bibinfo{author}{\bibfnamefont{B.}~\bibnamefont{Koslowski}},
  \bibinfo{journal}{J. Appl. Phys.} \textbf{\bibinfo{volume}{94}},
  \bibinfo{pages}{1478} (\bibinfo{year}{2003}).

\end{thebibliography}

\end{document}